\documentstyle[11pt,moriond,epsfig]{article}

\def\Journal#1#2#3#4{{#1} {\bf #2}, #3 (#4)}

\def\PLB{{\em Phys. Lett.}  B}

\def\PRD{{\em Phys. Rev.} D}

\def\be{\begin{equation}}
\def\ee{\end{equation}}
\def\bea{\begin{eqnarray}}
\def\eea{\end{eqnarray}}

\begin{document}
\vspace*{4cm}
\title{Measurement of the z- and x-dependence of the Mean Transverse
Momentum \\
of Charged Hadrons, Charged Pions and $K_s^0$ at HERMES}
\author{ A.A.Jgoun (on behalf of the HERMES collaboration)} 
\address{HEPD, PNPI, St.Petersburg, Russia}
\maketitle
\abstracts{
The mean transverse momenta of hadrons, $<P_{\bot}>$, produced
in $e^+N$ deep inelastic scattering was studied
 as a function of the hadron energy fraction $z$ and the Bjorken variable $x$. 
The statistical accuracy of the  HERMES data permitted to observe at high $z$
differences in $<P_{\bot}>$ between hadrons with
positive and negative charges, $h^+$, $h^-$ and $K_s^0$ for the first time.
 These differences can presumably be explained by the flavour dependence of the
intrinsic transverse momentum distributions of quarks in the initial
nucleon.}

\section{Introduction}

The mean transverse momentum $<P_{\bot}>$  of hadrons $h$
produced in semi-inclusive deep inelastic scattering (SIDIS) 
$e+N\to e^{\prime}+h+X$, studied in dependence on kinematic variables,
can give information about the quark transverse momentum distribution in the 
initial nucleon and the hadronization process. High-$z$ data are  more sensitive to 
the transverse momentum distribution of quarks in the initial nucleon whereas 
low-$z$ data are preferentially influenced by the quark fragmentation process. 
The flavour dependence
of the primordial momentum distribution of initial quarks can be investigated by 
studying $<P_{\bot}>$ for different hadrons produced in SIDIS.
The flavour dependence of the internal quark momentum distribution can manifest 
itself as a target isospin dependence of $<P_{\bot}>$ and can be studied in 
SIDIS on targets with different isospin.

\section{Results}

The cross section of hadron production in SIDIS is determined by five
independent variables $Q^2,\;x,\;z,\;P_{\bot},\;\phi^{\gamma}$. Here $-Q^2$ is 
the  virtual photon four-momentum squared, $x$ is the Bjorken scaling 
variable, $z$ is the ratio of the hadron energy to the photon energy in the 
laboratory system, and $P_{\bot}=|\vec{P}_{\bot}|$ is the transverse momentum of 
the hadron $h$ with respect to the direction of the photon. The azimuthal angle 
$\phi^{\gamma}$ is the angle between the positron scattering plane and  the hadron 
production plane, both having the direction of the virtual photon in common.

The analysis of the experimental data was based on a multi-dimensional 
unfolding procedure. The radiative corrections to $P_{\bot}$ and to the number of 
events observed in every bin
have been calculated with the programs GMC and POLRAD2~\cite{POLRAD2}. 
The main systematic uncertainties are those related to the unfolding procedure and
they were estimated to be less than $1.5\%$ for $h^\pm-$ production. The systematic 
uncertainty due to the
$\phi^{\gamma}$-dependence of the cross section and the detector efficiency is
 less than $1\%$. Other uncertainties (particle misidentification, 
cut dependence of the results etc.) are smaller compared to these numbers. For $K_s^0$ production the total
systematic uncertainty is less than $3\%$.

All systematic uncertainties have been added in quadrature to the statistical 
errors, the combined values are shown in the figures.

\vspace{-0.5cm}
\subsection{Dependence of $<P_{\bot}>$ on z}

\begin{figure}
  \epsfig{file=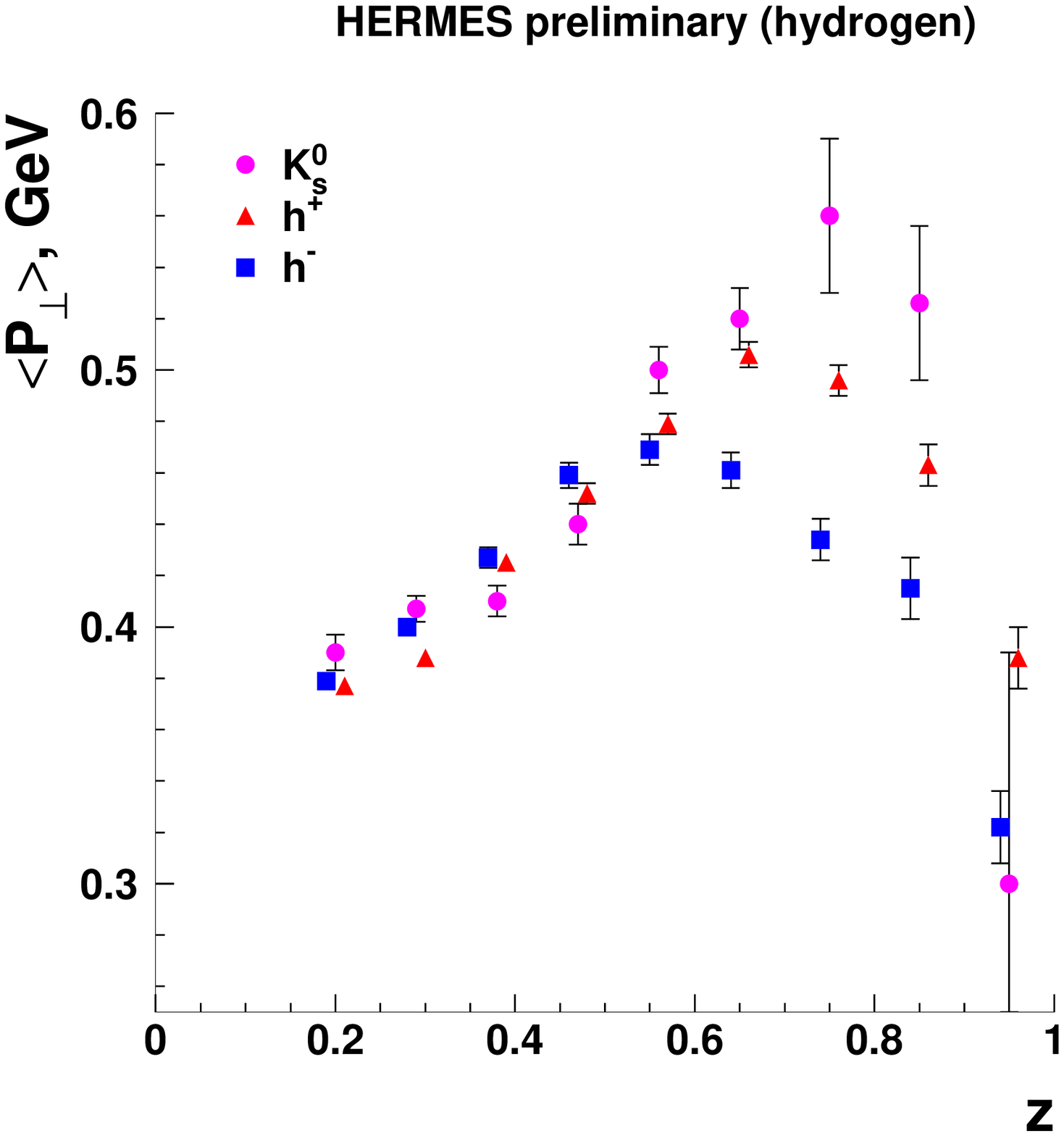, width =7.5cm,height=7cm}
  \epsfig{file=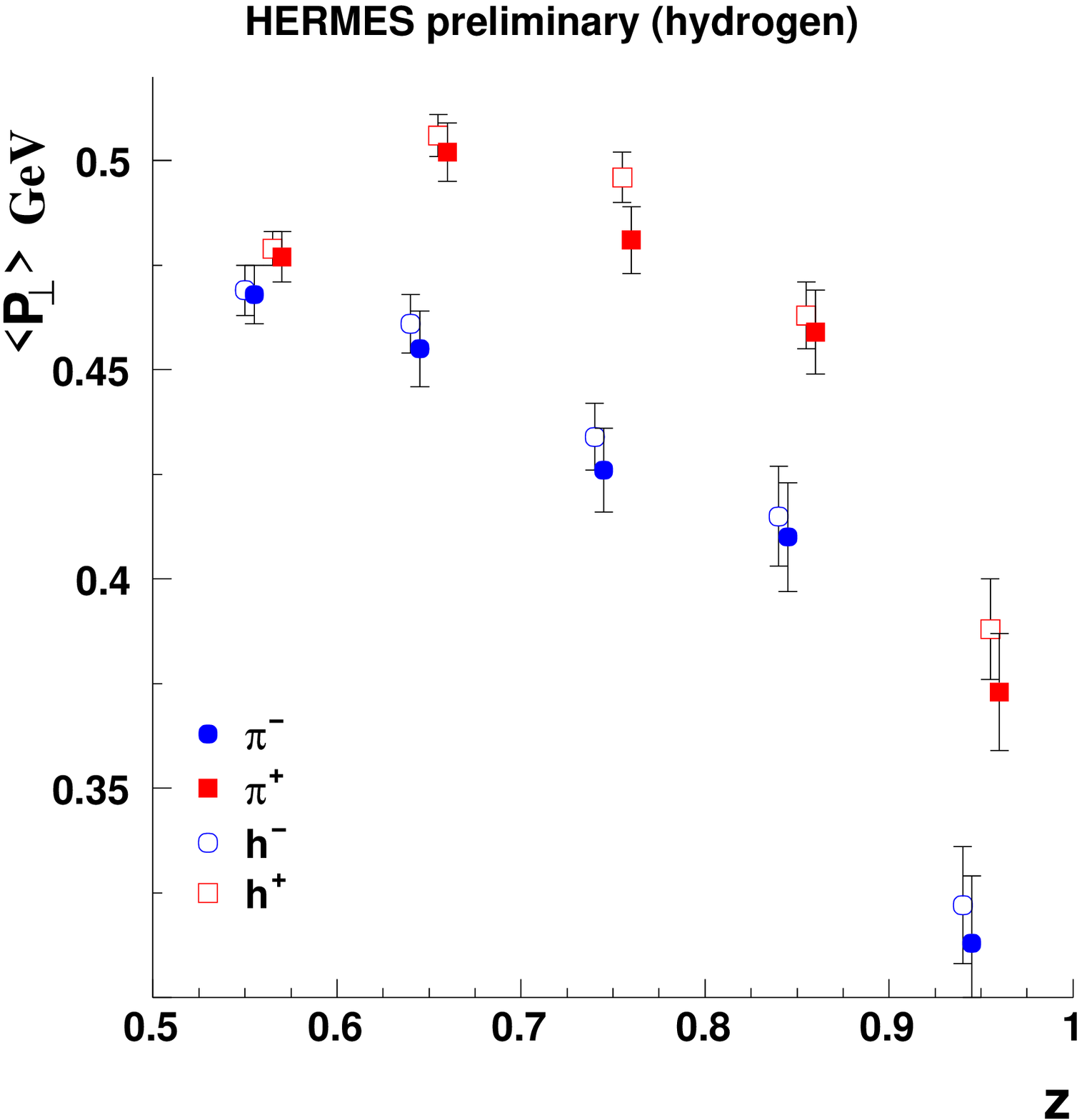, width =7.5cm,height=7cm}
  \vspace{-0.3cm}
  \caption{\it Behaviour of $<P_{\bot}>$ for $K_s^0$ and $h^\pm$ over the full
               measured $z$-range (left). Comparison of $h^\pm$ and $\pi^\pm$ data
               in the high-$z$ range (right).}
  \label{pt_mean}
  \begin{center}
    \rule{10cm}{0.2mm}
  \end{center}
\vspace{-0.5cm}
\end{figure}

From the left panel of Fig.~\ref{pt_mean} the $z$-dependences of 
$<P_{\bot}>$ for $h^+,\;h^-$ and $K_s^0$ can be seen to be similar for $0.2<z<0.6$. 
In this region $<P_{\bot}>$ increases approximately linearly with $z$. 
The main part of the statistics is located in this region of $z$ so that
the slopes of the cross sections versus $P_{\bot}^2$ are approximately equal to each 
other for all particles considered above. 

For $z>0.6$ HERMES data show a significant difference in the dependence of $<P_{\bot}>$
on z for $h^+,\;h^-$ and $K_s^0$: the average transverse momenta of $h^+$ and $K_s^0$
are higher than that for $h^-$. It appears worth noting  that some indication of this 
effect exists in the SLAC~\cite{ep} and EMC~\cite{emc} data already, albeit the large 
statistical uncertainties precluded a clear statement. 

\begin{figure}[ht]
\vspace{-0.3cm}
  \epsfig{file=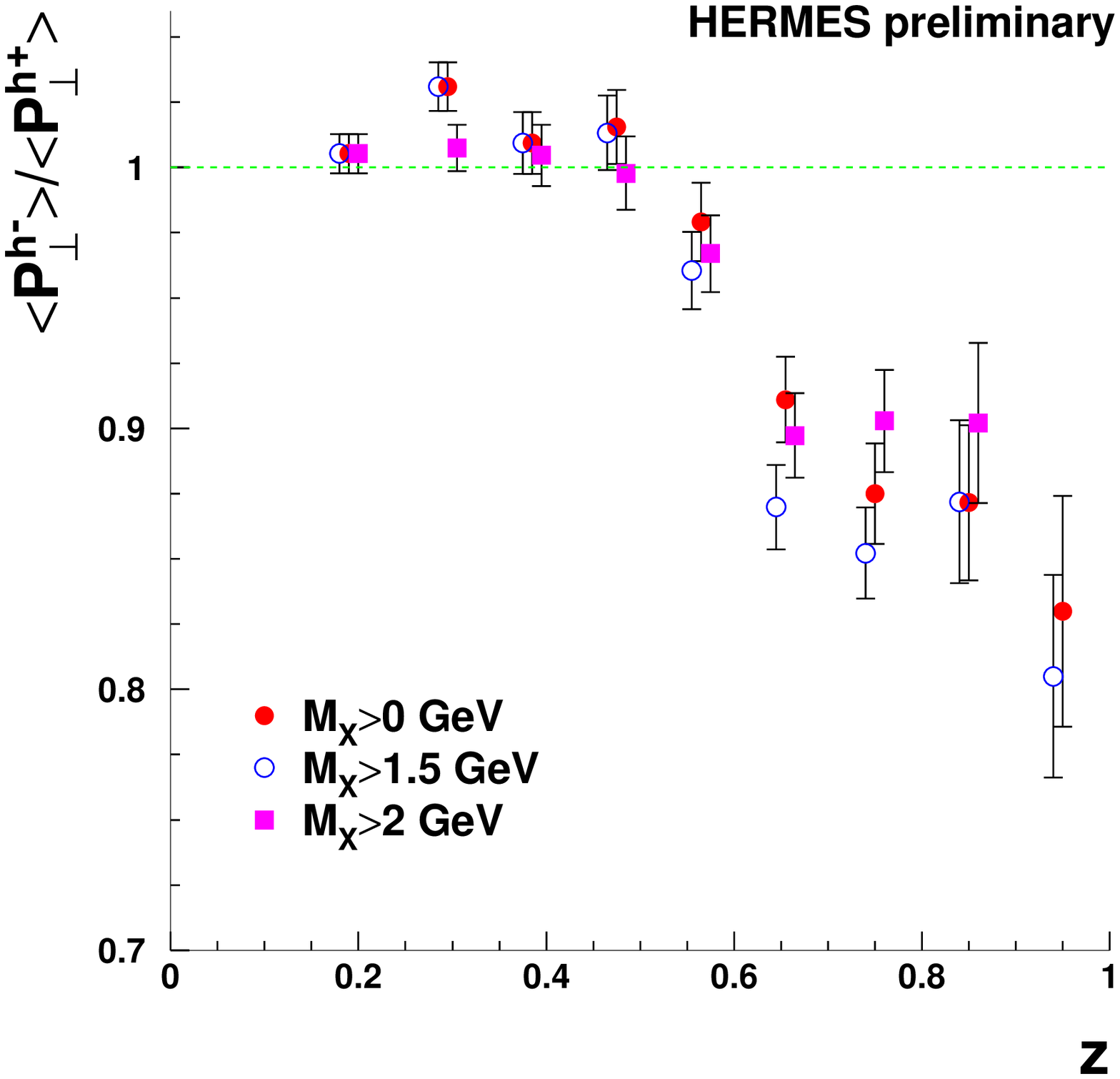, width =7.5cm,height=7.5cm}
  \epsfig{file=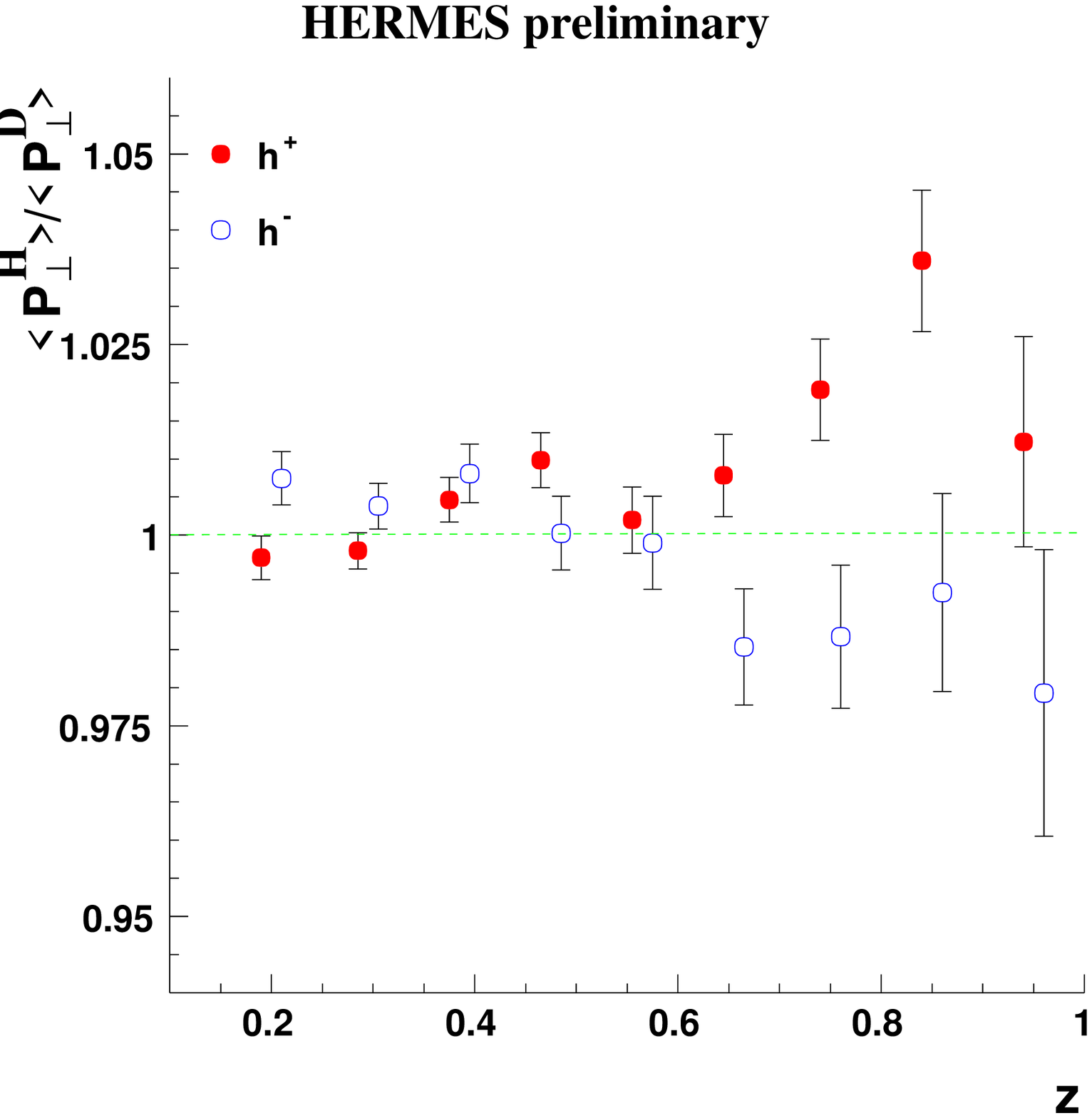, width =7.5cm,height=7cm}
 \vspace{-0.3cm}
 \caption{\it Ratios of $<P_{\bot}>$ for $h^-$ and $h^+$ vs. $z$ for different cuts 
  on $M_X$ (left) and ratios of $<P_{\bot}>$ for hydrogen and deuterium targets vs. $z$
  for $h^+$ and $h^-$ (right).}
  \label{mx_ratio}
  \begin{center}
    \rule{10cm}{0.2mm}
  \end{center}
\vspace{-0.5cm}
\end{figure}

The pion fraction in the hadronic events is about 80-90$\%$ for $z> 0.6$. 
From Fig.~\ref{pt_mean} (right) can be seen that within statistical uncertainties
$<P_{\bot}>$ for pion and hadron agrees with one another, for both charges.
%
In principle, the difference between $\pi^+$ and $\pi^-$ can be, at least partially,
caused by exclusive processes. However, Fig.~\ref{mx_ratio} (left) shows that the 
application of different cuts in the missing mass up to 2 GeV does not change the 
observed difference of $<P_{\bot}>$ between $\pi^+$ and $\pi^-$.

The difference of $<P_{\bot}>$ between $\pi^+$ and $\pi^-$ could possibly be related to
the fragmentation process. Indeed, usually the contribution of $u$-quarks dominates in 
deep inelastic lepton-nucleon scattering and $u$-quarks can create $\pi^+$-mesons in a
one-step process picking up $\bar{d}$-quarks from the vacuum. In contrast, a $u$-quark 
producing a $\pi^-$ needs at least a three-step process: the $u$-quark radiates first
a gluon that subsequently converts into a $d\bar{d}$-pair from which the $d$-quark then
picks up a $\bar{u}$-quark from the vacuum. The transverse momentum acquired in such a
three-stage process is on average greater than that in the one-step process, 
while $Q^2$, $y$ and $z$ remain unaffected. Hence the transverse momenta of $\pi^-$ 
should be higher than that of $\pi^+$ at large $z$, where few-step processes are more 
important than at low $z$ where multi-step processes dominate. While Monte Carlo
simulations confirm these qualitative considerations, the experimental data are in 
the strong contradiction with this picture; $<P_{\bot}>$ for $\pi^+$ is measured to be
greater than that for $\pi^-$. This presumably means that the difference of 
$<P_{\bot}>$ between $h^+$ and $h^-$ is not a result of some difference in their 
fragmentation.

The assumption that the transverse momentum distribution of quarks in the nucleon 
is strongly flavour-dependent would manifest itself in a difference of $<P_{\bot}>$ 
for differently charged leading mesons in SIDIS at high $z$. If then the difference 
in $<P_{\bot}>$ between $\pi^+$ and $\pi^-$ is related to the difference of the 
internal transverse momenta of $u$- and $d$-quarks, a dependence of $<P_{\bot}>$ 
on the target isospin may be anticipated. Indeed, the HERMES data indicate that such 
a tendency might exist, as can be seen from Fig.~\ref{mx_ratio} (right). 
The following MC parameters were tuned to arrive at a decent description of the data: 
the average intrinsic momenta 
for $u$- and $d$-quarks, average fragmentation transverse momentum and minimal mass 
of the string system.  Neither $z$- nor $x-$dependences of $<P_{\bot}>$ could be 
described satisfactorily in the full kinematic region and the obtained  parameters for 
$u$- and $d$-quarks ($<q^u_{\bot}>=0.54$ GeV and $<q^d_{\bot}>=0.16$ GeV) are so 
much different from one another that they appear unrealistic.

\subsection{Dependence of $<P_{\bot}>$ on x}

\begin{figure}[ht]
  \epsfig{file=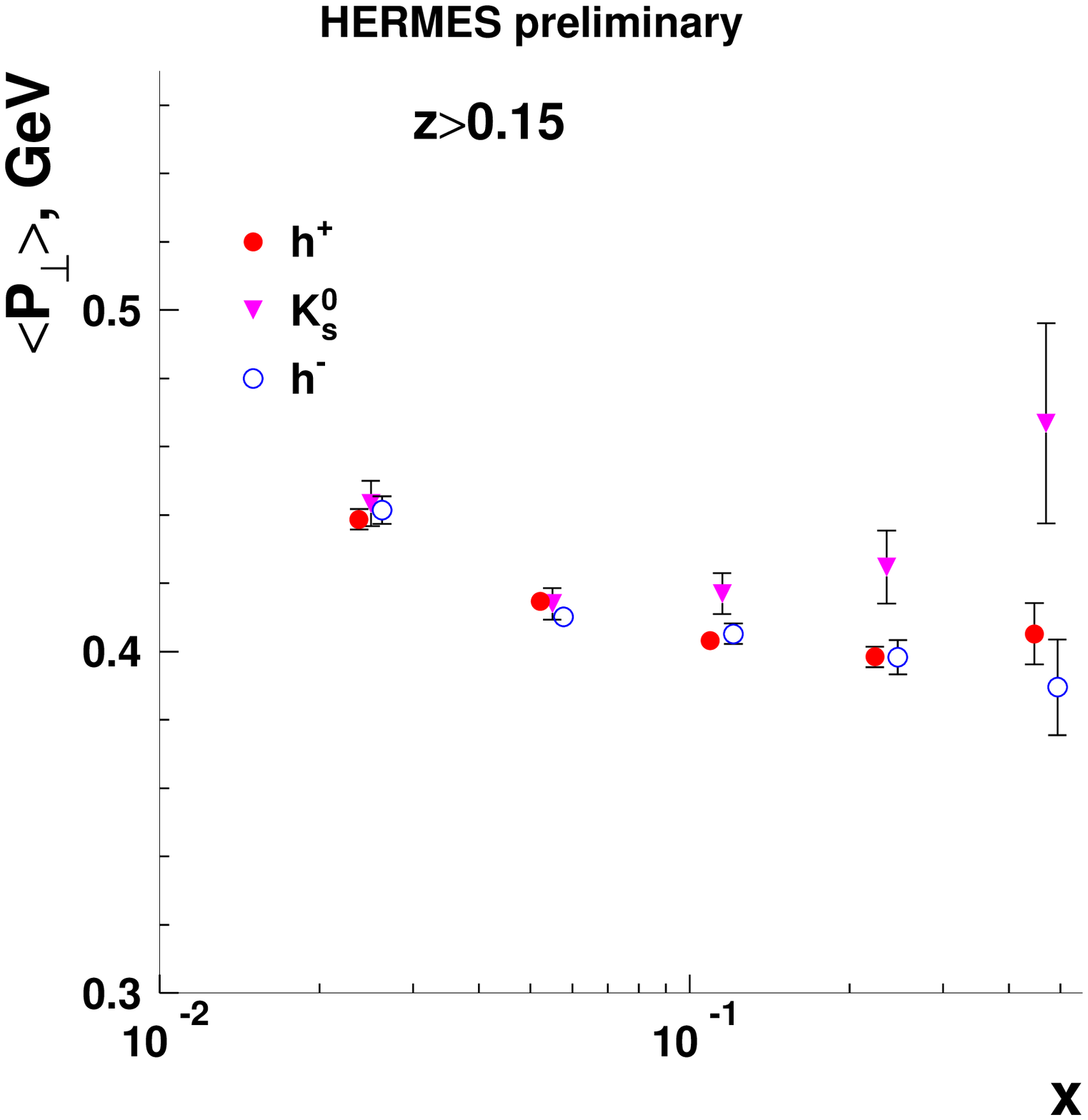, width =7.5cm,height=7cm}
  \epsfig{file=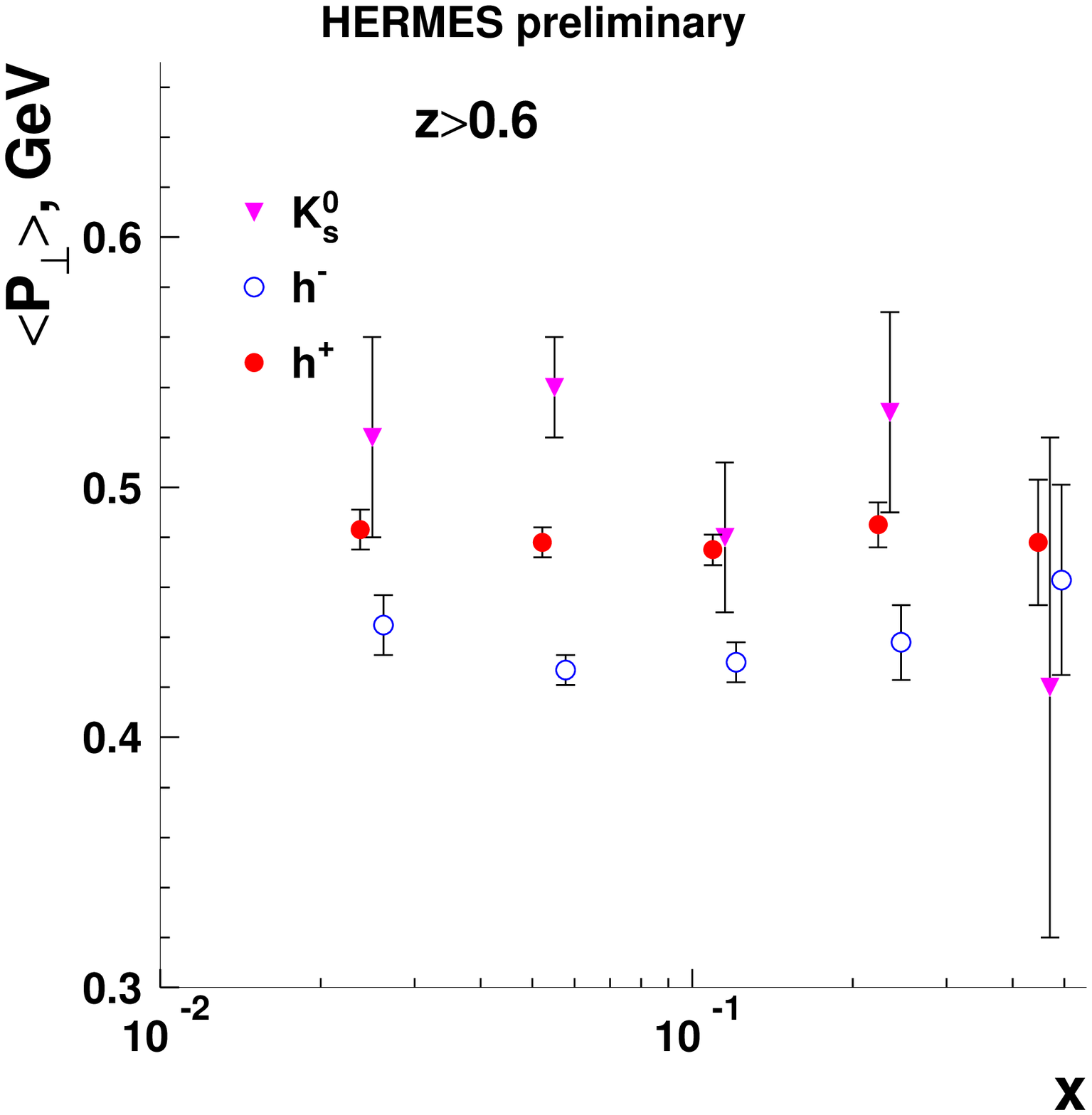, width =7.5cm,height=7cm}
  \vspace{-0.3cm}
  \caption{\it $<P_{\bot}>$ for $K_s^0$ and $h^\pm$ vs. $x$ for a hydrogen target, 
               shown for $z>0.15$ (left) and $z>0.6$ (right).}
  \label{x_dep} 
  \begin{center}
    \rule{10cm}{0.2mm}
  \end{center}
  \vspace{-0.7cm}
\end{figure}

Figure~\ref{x_dep} shows the $x$-dependence of $<P_{\bot}>$ for $h^+$, $h^-$ and $K_S^0$ 
produced on hydrogen for different $z$-cuts. As can be seen from Fig.~\ref{x_dep} 
(left), the $x$-dependences of $<P_{\bot}>$ for $h^+$ and $h^-$ coincide within the
statistical uncertainty.
Applying instead the cut $z> 0.6$, the relative contribution of the primordial 
transverse  
momentum of the initial quarks to $<P_{\bot}>$ is increased. A clear difference 
of  $<P_{\bot}>$ between $h^-$, $h^+$ and $K_s^0$ appears at large $z$, see 
Fig.~\ref{x_dep} (right). No significant $x$-dependence of $<P_{\bot}>$ can be observed
in the region $z >  0.6$,
this agrees well with the factorization property of the transverse and longitudinal momentum distributions of quarks in the nucleon.

\section{Conclusions}

The main features of the behaviour of average transverse momenta as seen in the 
HERMES data are the following:
\begin{itemize}
\item Similar behaviour of average transverse momenta of  
      $K^0_s$, $h^+$, and $h^-$ at $z<0.5$.
\item A clear difference of $<P_{\bot}>$ between positive and negative hadrons 
      for $z>0.6$.
\item The LUND model of fragmentation predicts a higher transverse momentum for $h^-$
      than for $h^+$ at $z>0.6$, while HERMES data show the opposite.
\item The difference in $<P_{\bot}>$ for $h^+$, $h^-$ and $K_s^0$ can be explained 
      by the hypothesis that the internal transverse momentum distributions of quarks 
      in the nucleon are flavour dependent. The difference of $<P_{\bot}>$ between 
      $h^+$ 
      and $h^-$ observed in DIS on hydrogen and deuterium targets does not contradict 
      this hypothesis.
\end{itemize}

{\bf Acknowledgement}

I would like to thank all my colleagues at my home institute (Drs. S.I. Manaenkov, G.Gavrilov) and at DESY (Prof. W.-D. Nowak, drs. E.-C. Aschenauer and  M. Vincter) who have helped in the preparetion of this work.

\end{document}